\documentclass[aps,prb,twocolumn,showpacs]{revtex4}

\usepackage{mathptm}    
\usepackage{dcolumn}                    
\usepackage{bm}                         
\usepackage{graphicx}
\usepackage{amsmath}

\usepackage{appendix}
\usepackage{subfigure}

\usepackage{times}

\usepackage{natbib}

\begin{document}

\title{Possible pairing states of the Fe-based superconductors}

\author{Yunkyu Bang}
\email[Author to whom the correspondences should be addressed:
]{ykbang@chonnam.ac.kr} \affiliation{Department of Physics,
Chonnam National University, Kwangju 500-757, and Asia Pacific
Center for Theoretical Physics, Pohang 790-784, Korea}

\author{Han-Yong Choi}
\affiliation{Department of Physics and Institute for Basic Science
Research, SungKyunKwan University, Suwon 440-746, Korea}

\begin{abstract}
We consider the minimal two-band model for the Fe-based
superconductors with a phenomenological pairing interaction which
mimics short-range antiferromagnetic (AFM) fluctuations. Two
superconducting (SC) gap solutions are found to exist with the
model: sign-changing s-wave gap ($\pm$s-wave) and double d-wave
gap states. Both solutions hold the approximate relation
$\Delta_{h} ^{max} N_h \approx \Delta_{e} ^{max} N_e$, a generic
feature of two band model with a dominant interband pairing
interaction. We carried out the calculations of the SC properties
of the both SC states such as the density of states, temperature
dependencies of spin-lattice relaxation rate $1/T_1$, Knight
shift, and penetration depth, particularly taking into account of
the interband coherence factors. The results are discussed in
comparison with the currently available experimental data.

\end{abstract}


\date{\today}
\maketitle

\section{Introduction}

Recent discovery of the Fe-based superconducting compounds
provided a great impetus to the research of superconductivity
(SC). Since the first report on the superconducting transition at
7K with the doped LaOFeP by Kamihara et al. \cite{Kamihara},
various substitutions (mainly, P by As and La by Ce, Gd, Sm, Pr)
for this mother compound increase the superconducting temperature
$T_c$ over 50K with Sm(OF)FeAs \cite{Chen}. Intensive
investigations by many experimentalists and theorists have already
revealed main metallic and superconducting properties of this
group of materials.

From band calculations\cite{Singh,Haule,Mazin1,Mazin2,Cao,bands},
it is agreed on that the 3d electrons of Fe atoms are the main
contributors to the conduction bands crossing the Fermi surface
(FS). Besides the degree of degeneracy, the key feature of
conduction bands is that it consist of hole band(s) around
$\Gamma$ point and electron band(s) around $M$ point [in the
notation of the folded Brillouin-zone (BZ)
scheme\cite{Singh,Mazin1}].

Regarding the pairing symmetry, there are already many
experiments: (1) Knight shift below $T_c$ shows a clear drop
indicating a spin singlet pairing \cite{Curro}; (2) tunneling
spectroscopy \cite{tunneling} showed the zero-bias conductance
peak (ZBCP) - signature of a sign changing gap, but the
interpretation of the shape of density of states (DOS) $N(\omega)$
is diverse; (3) nuclear spin-lattice relaxation rate $1/T_1$
\cite{Curro,T1} unanimously showed no coherent peak and $\sim T^3$
dependence below $T_c$, hence strongly suggesting a d-wave type
gap; and (4) specific-heat coefficient $C(T)/T$ below $T_c$ (Ref.
\cite{C(T)}) -- although the measurement is not yet reaching low
enough temperature -- appears $T$ linear indicating the gap with
lines of node. All these experiments appear to be consistent with
a d-wave type gap. However, recent penetration depth measurements
with PrFeAsO, Sm(OF)FeAs and  Nd(OF)FeAs (Ref.\cite{pene})
strongly suggest a fully opened gap at low temperatures indicating
a s-wave type pairing symmetry.

Regarding the paring glues, the phonon interaction appears
unlikely mainly because the electron-phonon coupling is estimated
to be very weak ($\lambda < 0.2$)\cite{phonon}. On the other hand,
this series of materials, without doping, commonly has a spin
density wave (SDW) transition at around $\sim 150$ K. When the
superconductivity appears with doping, the SDW correlation is
expected to remain, albeit the long range order disappears.

Recent neutron-scattering experiments with La(OF)FeAs and
Ce(OF)FeAs (Ref. \cite{neutron}) directly measured the
antiferromagnetic (AFM)-type correlation of the Fe d-electron spin
moment. The overall phase diagram with doping for Ce(OF)FeAs
reveals a close correlation with an antiferromagnetism and
superconductivity, suggesting the important role of magnetic
fluctuations as a pairing glue. It also shows that the generic
phase diagram of these compounds shares the universal features
with the high-$T_c$ cuprates, Pu-115 superconductor \cite{Pu115},
and various heavy fermion superconductors; namely, the SC occurs
in the neighborhood of the magnetic long range order when this
magnetic order is suppressed.  In particular, the magnetic order
is an AFM type. This universal phase diagram is very tantalizing
because it appears to cover a wide class of unconventional SC
materials with a range of $T_c$ from a few mK to 100 K and
suggests that the AFM fluctuation is a common thread and its
characteristic energy roughly scales with the SC $T_c$
\cite{Pu115}.

For the Fe-based SC materials, several theoretical models were
already proposed and most of them started with the orbital basis
of the Fe 3d-electrons including Hubbard $U$ interaction(s) and
Hund coupling(s) $J$ \cite{Mazin1,Eremin,DHLee,Kuroki,others}.
Some of these studies \cite{Mazin1,Eremin,DHLee} found that the
$\pm$s-wave gap as a dominant instability.  A d-wave gap also
often appears as a second instability \cite{DHLee,Kuroki}. In this
paper, we took a phenomenological approach to investigate possible
pairing states in the Fe-pnitide superconductors. The
non-interacting part of Hamiltonian is constructed by choosing a
minimal set of topologically distinct two bands and the
interaction part of Hamiltonian is assumed from the experimental
input \cite{neutron,resonance}, simulating a short range AFM
correlation. By solving the coupled BCS gap equations, we found
the two SC gap solutions: a sign-changing s-wave gap and a double
d-wave gap. For the both SC states, we carried out the
calculations of the SC properties such as the DOS, temperature
dependencies of the spin-lattice relaxation rate $1/T_1$, Knight
shift, and penetration depth. We particularly take into account of
the interband coherence factors, unique to the two band model, in
these calculations. The results are discussed in comparison with
the currently available experimental data.

\begin{figure}
\noindent
\includegraphics[width=80mm]{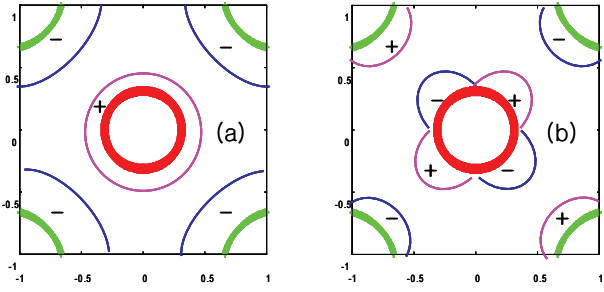}
\caption{(Color online) FSs of $\epsilon_{h} (k)$ (red) and
$\epsilon_{e} (k)$ (green) bands and two gap solutions considered
in the paper: (A) $\pm$s-wave gap and (B) double d-wave gap. The
width of the FS represents the local DOS for each band within
$\omega_{AFM}=20 meV$ energy.  \label{fig1}}
\end{figure}
\section{Model}

We propose a minimal phenomenological two band model for the
Fe-based superconductors. For the non-interacting part of
Hamiltonian, we observe that several band structure calculations
\cite{Singh,Haule,Mazin1,Mazin2,Cao,bands} of the Fe-pnictide
compounds reached the consensus that the FS of the doped compounds
consists of two hole pockets and two electron pockets. To keep the
essential physics, but avoiding unnecessary complexity, we choose
only two topologically distinct bands: one hole band around
$\Gamma$ point (0,0) and one electron band around $M$ point ($\pm
\pi,\pm \pi$).

The main phenomenological assumption of our model is the
interacting part of Hamiltonian. This pairing interaction $V({\bf
q})$ is chosen to simulate a short range AFM spin fluctuations
peaking at the ordering wave vector ${\bf Q}=(\pm \pi,\pm \pi)$.
This assumption is directly motivated by the experimental
observations of the AFM correlation in the Fe-pnictide compounds
by the neutron-scattering measurements \cite{neutron,resonance}.
The elastic neutron-scattering experiments for
La(F$_x$O$_{1-x}$)FeAs and Ce(F$_x$O$_{1-x}$)FeAs by Dai and
co-workers \cite{neutron} showed the long-range AFM order of the
Fe3d-electron spins for the doping range of $x=0 - 0.05$. When
this long-range AFM order disappears beyond the doping around
x=0.05, SC appears up to the doping range x $\sim$ 0.2 (this is
only the limit of the measured data). This overall phase diagram
appears to be generic for the Fe-pnictides, and clearly shows that
the AFM correlation is the dominant magnetic correlation in this
group of materials.

More importantly, this AFM correlation is expected to continue to
exist in the doping range where the SC phase occurs albeit
becoming a short-rang one. This speculation is supported by more
recent inelastic neutron-scattering experiment in the doped
(BaK)Fe$_2$As$_2$ compound ($T_c$=38K)  by Christianson et al.
\cite{resonance} In this experiment, a clear magnetic-resonance
peak is observed at the expected position of the AFM correlation,
i.e., at ($\pi,\pi$) momentum. With this series of experiments it
is clear that the AFM correlation is the dominant magnetic
correlation in the Fe-pnictides among other competing magnetic
correlations such as weak ferromagnetism \cite{Singh,Haule},
checkerboard AFM \cite{Mazin2,Cao}, and AFM stripe phase
\cite{Mazin2,stripe}, which were theoretically proposed. Finally,
the coupling matrix element is assumed to be a constant for
simplicity.  The Hamiltonian is written as

\begin{eqnarray}
H &=& \sum_{k \sigma} \epsilon_h (k) h^{\dag}_{k \sigma} h_{k
\sigma} + \sum_{k \sigma} \epsilon_e (k) e^{\dag}_{k \sigma} e_{k
\sigma} \nonumber \\
& & +\sum_{k k^{'} \uparrow \downarrow} V(k,k^{'}) h^{\dag}_{k
\uparrow} h^{\dag}_{-k \downarrow}
h_{k^{'} \downarrow}h_{-k^{'} \uparrow} \nonumber \\
& & +\sum_{k k^{'} \uparrow \downarrow} V(k,k^{'}) e^{\dag}_{k
\uparrow} e^{\dag}_{-k \downarrow} e_{k^{'} \downarrow}e_{-k^{'}
\uparrow} \nonumber \\
& & + \sum_{k k^{'} \uparrow \downarrow} V (k,k^{'}) h^{\dag}_{k
\uparrow} h^{\dag}_{-k \downarrow} e_{k^{'} \downarrow}e_{-k^{'}
\uparrow} \nonumber \\
& & +\sum_{k k^{'} \uparrow \downarrow} V(k,k^{'}) e^{\dag}_{k
\uparrow} e^{\dag}_{-k \downarrow} h_{k^{'} \downarrow}h_{-k^{'}
\uparrow}
\end{eqnarray}

\noindent where $h^{\dag}_{k \sigma}$ and $e^{\dag}_{k \sigma}$
are the electron creation operators on the hole and the electron
bands, respectively. $\epsilon_{h,e} (k)$ are the dispersions of
the hole band and electron bands, respectively, defined as
$\epsilon_{h} (k)=t_1 ^h (\cos k_x +\cos k_y) + t_2 ^h \cos k_x
\cos k_y + \epsilon^h$ and $\epsilon_{e} (k)=t_1 ^e (\cos k_x
+\cos k_y) + t_2 ^e \cos \frac{k_x}{2} \cos \frac{k_y}{2} +
\epsilon^e$. In this paper, we choose the band parameters as
(0.30,0.24,-0.6) for hole band and (1.14,0.74,1.70) for electron
band with the notation ($t_1, t_2, \epsilon$) \cite{Eremin}.

The pairing interaction $V(k,k^{'})$ is phenomenologically defined
below. It is all repulsive in momentum space and it represents a
short range AFM spin fluctuations as explained above.

\begin{equation}
V(k,k^{'}) = V_M \frac{\kappa^2}{|(\vec{k}-\vec{k^{'}})-\vec{Q}|^2
+\kappa^2}
\end{equation}

\noindent where $\vec{k}$ and $\vec{k^{'}}$ are the two-
dimensional momenta on the two dimensional BZ and the parameter
$\kappa$ controls the magnetic correlation length as $\xi_{AFM} =
2 \pi a/ \kappa$ ($a$ is the unit-cell distance). This interaction
mediates the strongest repulsion when two momenta $\vec{k}$ and
$\vec{k^{'}}$ are spanned by the ordering wave vector $\vec{Q}$.
This condition is better fulfilled when the two momenta $\vec{k}$
and $\vec{k^{'}}$ reside each other on different bands  in the
model band structure (see Fig.1). As a result, the sign-changing
s-wave gap can form on each band as already suggested by several
papers \cite{Mazin1,Eremin,DHLee,Kuroki}. However, this opposite
sign gap on the hole and electron bands is not limited with the
$\pm$s-wave state ([Fig.1 (a)]. Another possibility, which
conforms to the lattice symmetry, is that each band develops a
d-wave gap but with $\pi$ phase shift between two bands [Fig.1
(b)]. We call this type of gap as double d-wave gap.

We need to mention that our model did not include the screened
Coulomb interaction (neither did the other theoretical
investigations \cite{Mazin1,Eremin,DHLee,Kuroki}), which certainly
exists in the Fe-pnitide superconductors as well as in all metals
in general.
The screened Coulomb interaction is traditionally treated as
"Coulomb pseudo potential" $\mu^*$ in the conventional phonon-
driven SC. However, the reliable estimate for its strength is
practically impossible because even a small difference in $\mu^*$
would cause a large change in $T_c$. In the Fe-pnictides, if we
are to determine $T_c$ theoretically, a quantitative estimate of
$\mu^*$ is necessary. We did not include it in our model
interaction, first, because we do not know how to reliably
estimate it in these compounds and, second, because the primary
purpose of the present paper is not the prediction of the precise
$T_c$. Still we could investigate its generic effects on the
different pairing symmetries such as $\pm$ s-wave and double
d-wave gaps; for example, how large value of $\mu^*$ is necessary
to kill the $\pm$ s-wave pairing for a given strength of the AFM
interaction. We think that this kind of analysis will dilute the
focus of the present paper and therefore should be a separate
investigation.
We briefly remark, however, on the general effects of the screened
Coulomb interaction. The screened Coulomb interaction becomes a
short-range interaction in real space and therefore weakly
momentum dependent in momentum space. This type of interaction is
almost harmless for the d-wave type pairing but extremely
detrimental for the s-wave type pairing. The pairing solutions in
this paper should be considered with this point in mind.

Now we solve the Hamiltonian Eq.(1) using the BCS approximation
and the two band electrons need two SC order parameters (OPs)

\begin{eqnarray}
\Delta_{h} (k) & = &  \sum_{k^{'} } V(k,k^{'}) <h_{k^{'} \downarrow} h_{-k^{'} \uparrow}> , \\
\Delta_{e} (k) & = &  \sum_{k^{'} } V(k,k^{'}) <e_{k^{'}
\downarrow} e_{-k^{'} \uparrow}>.
\end{eqnarray}

\noindent After decoupling the interaction terms of Eq.(1) using
the above OPs, the self-consistent mean field conditions lead to
the following two coupled gap equations.

\begin{eqnarray}
\Delta_h (k)  &=&  \\ \nonumber  - & \sum_{k^{'} } & [V_{hh}
(k,k^{'}) \Delta_h (k^{'})  \chi_h (k^{'}) + V_{he}
(k,k^{'})\Delta_e (k^{'})  \chi_e (k^{'})], \\ \nonumber
\Delta_e (k)  &=&  \\ \nonumber  - & \sum_{k^{'} }& [V_{eh}
(k,k^{'}) \Delta_h (k^{'})  \chi_h (k^{'}) + V_{ee}
(k,k^{'})\Delta_e (k^{'})  \chi_e (k^{'})]. \\
\end{eqnarray}
\noindent where $V_{hh} (k,k^{'})$, $V_{he} (k,k^{'})$, etc are
the same interaction defined in Eq.(2) but the subscripts are
written to clarify the meaning of $V_{hh} (k,k^{'})$ =$V
(k_h,k^{'} _h)$, $V_{he} (k,k^{'})$ =$V (k_h ,k^{'}_e)$, etc., and
$k_h$  and $k_e$ specify the momentum $k$  located on the hole and
electron bands, respectively.
The pair susceptibilities are defined as
\begin{eqnarray}
\chi_{h,e}(k) &=& N(0)_{h,e} \int _0 ^{\omega_{AFM}} d \xi
\frac{\tanh (\frac{E_{h,e}(k)}{2 T})}{E_{h,e} (k)}
\end{eqnarray}

\noindent where $E_{h,e} (k) =\sqrt{\xi^2 + \Delta_{h, e}^2 (k)}$,
and $N(0)_{h,e} $ are the quasiparticle excitations and the DOS of
the hole and electron bands, respectively, and $\omega_{AFM}$ is
the cutoff energy of the pairing potential $V(q)$.

When we solve the above gap equations Eq(5) and Eq.(6), we
numerically restricted the momenta $k_{h,e}$ and $k^{'}_{h,e}$
around the FSs of the hole and electron bands within
$\omega_{AFM}$ energy range. Therefore, the FS shapes and the
local DOS $N(0)_{h,e}$ of the realistic bands are faithfully taken
into account in our gap solutions. Also no restriction on the
functional forms of the gaps $\Delta_{h,e} (k)$ was imposed except
the general symmetry depicted in Fig.1, so that the $k$-dependence
of the gap functions $\Delta_{h, e} (k)$ will follow the
characteristics of the bands and pairing interaction.

\section {Gap solutions}

As explained in Sec. II, the main pairing process with the AFM
spin-fluctuation mediated interaction $V(q)$ is the interband pair
hopping between the hole  and the electron bands, in which a pair
of electrons $(k,-k)$ on the hole band scatters to a pair of
electrons $(k^{'},-k^{'})$ on the electron band and vice versa.
This process is particularly dominant when the size of the FS of
each band is much smaller than the size of $\vec{Q}$ vector.
Considering only this interband pair process [keeping only
$V_{he}$ and $V_{eh}$ terms in Eqs.(5) and (6)], we observe the
fact that the pair potential $\Delta_h(k)$ for the hole band
electrons is provided by the pairs of electrons in the electron
band and vice versa. The physical consequence of it is that the
relative sizes of the gaps and DOSs on each band are reversed;
namely, if $N_h(0)
> N_e(0)$, then $|\Delta_h(k)| < |\Delta_e(k)|$ holds in general.
This relation holds both for the $\pm$s-wave and for the double
d-wave gap solutions and affects all superconducting properties
such as tunneling DOS, Knight shift, $1/T_1$, and penetration
depth.

For all numerical calculations in this paper, we choose the
parameters $\kappa=0.2 \pi$ ($ \xi_{AFM} \sim 10 a$),
$\omega_{AFM} =20 meV $, and $V_M=10 eV$ (average interaction
$<V(q)> =1.115 eV$). Our choice of band parameters produces
$N_h(0)=0.74/eV$ and $N_e(0)=0.285/eV$ so that $N_h(0)/N_e(0)
\approx 2.6$. We think that these numbers represent the Fe-based
SC materials but should not be taken too seriously; in particular
the pairing strength $V_M=10 eV$ is chosen freely for
demonstration.

\subsection {$\pm$s-wave gap}

This solution for the Fe-based SC is already proposed by several
authors \cite{Mazin1,Eremin,DHLee,Kuroki}  Here we demonstrate
that this solution is indeed realized with a simple
phenomenological interaction, which mimics an AFM spin
fluctuations, on the minimal two band model representing the
Fe-based SC compounds.

As we described above, the reversed relation between the magnitude
of the DOSs and the size of gaps holds more rigorously for the
s-wave case and we suggest an approximate relation $\Delta_{h}
^{max} N_h \approx \Delta_{e} ^{max} N_e$ (see the Appendix for
more detailed discussions).
This relation is a generic feature of the model. Therefore, given
a substantial difference of DOS between the hole and electron
bands (several band calculations
\cite{Singh,Haule,Mazin1,Mazin2,Cao,bands} indicate that this is
true for the Fe-based SC materials), at least two distinctively
different sizes of the SC gaps should be observed in various
experiments \cite{Ding}. In particular, because the band with a
larger DOS would dominate the physical properties but actually
holds a smaller gap, this feature will modify various SC
properties of the Fe-based SC in unorthodox manner, such as
$\Delta /T_c$ value, temperature dependencies of various SC
properties below $T_c$, and the responses to impurities.
\begin{figure}
\noindent
\includegraphics[width=80mm]{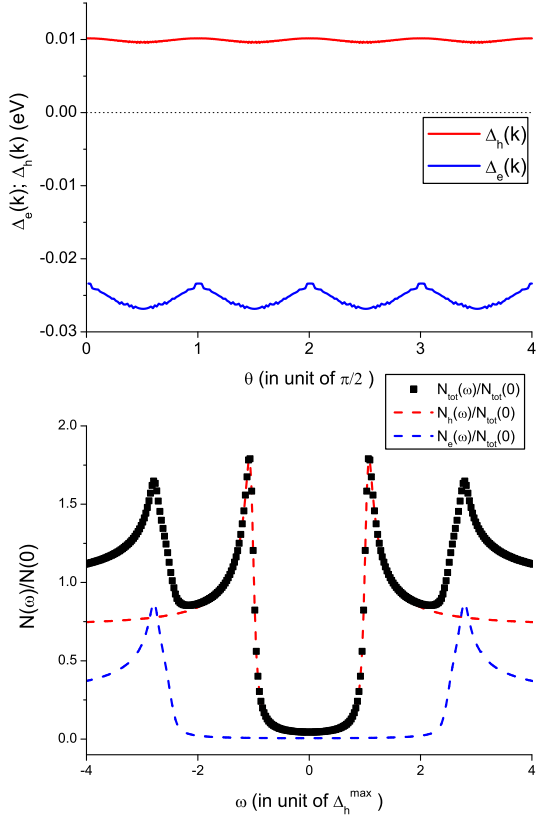}
\caption{(Color online) (a) $\pm$s-wave gap solutions $\Delta_h
(k)$ and $\Delta_e (k)$. (b) Normalized DOS of the hole band
N$_{h} (\omega)$ (red dotted line), electron band N$_{e} (\omega)$
(blue dotted line), and the total N$_{tot} (\omega)$ (solid black
squares). \label{fig2}}
\end{figure}

In Fig.2, the solution of the $\pm$s-wave gap and the
corresponding DOS are shown. As mentioned, $N_h(0)=2.6 N_e(0)$ for
our bands. Accordingly the size of gaps of the hole band and the
electron band is reversed as  $\Delta_h
 ^{max} \approx 10 meV$ and $\Delta_e ^{max} \approx 25 meV$.
The strongly momentum dependent pairing interaction and the
realistic bands naturally induce an anisotropic modulation of the
s-wave gaps with $C_4$ symmetry; the modulation is stronger for
the larger gap on the smaller DOS band (electron band around $M$
point). Compared to the case of the double d-wave solution, the
average size of the $\pm$s-wave gap is larger by a factor of $\sim
5$ with the same pairing potential. Therefore, unless some other
interactions are added, the ground state of our model is the
$\pm$s-wave SC state. This conclusion is already obtained by other
authors \cite{Mazin1,DHLee,Eremin,Kuroki} with different models
and approaches.
The separate and total DOSs plotted in Fig.2(b) show the main
features of the $\pm$s-wave gap: two peak structure, the large DOS
with a small gap and the small DOS with a large gap. The overall
shape of the total DOS is not very much revealing compared to the
current tunneling DOS measurements \cite{tunneling}. However, it
is too early to make a decisive conclusion with our calculations
without including Andreev scattering.  Also the ZBCP, the hallmark
of a d-wave gap and observed in experiments with the Fe-based
superconductors \cite{tunneling}, can equally be obtained with the
$\pm$s-wave gap state.

We consider nuclear spin-lattice relaxation rate $1/T_1$ for the
$\pm$s-wave gap. Several groups \cite{Curro,T1} have reported that
$1/T_1$ shows no coherence peak and the $T^3$ power law below
$T_c$, strongly suggesting an unconventional gap with lines of
node such as a d-wave gap. s-wave gap is known to have a
constructive coherent factors for $1/T_1$ to induce the coherence
peak over a temperature range below $T_c$. However, as Mazin et
al.\cite{Mazin1} envisaged, the sign-changing gaps between two
bands provide a destructive coherent factor for the interband
scattering which will largely cancel the intraband coherent
factors. As a result the coherent peak of $1/T_1$ for the
$\pm$s-wave gap SC will be substantially reduced. The explicit
formula that we used for the calculations is the following:

\begin{figure}
\noindent
\includegraphics[width=90mm]{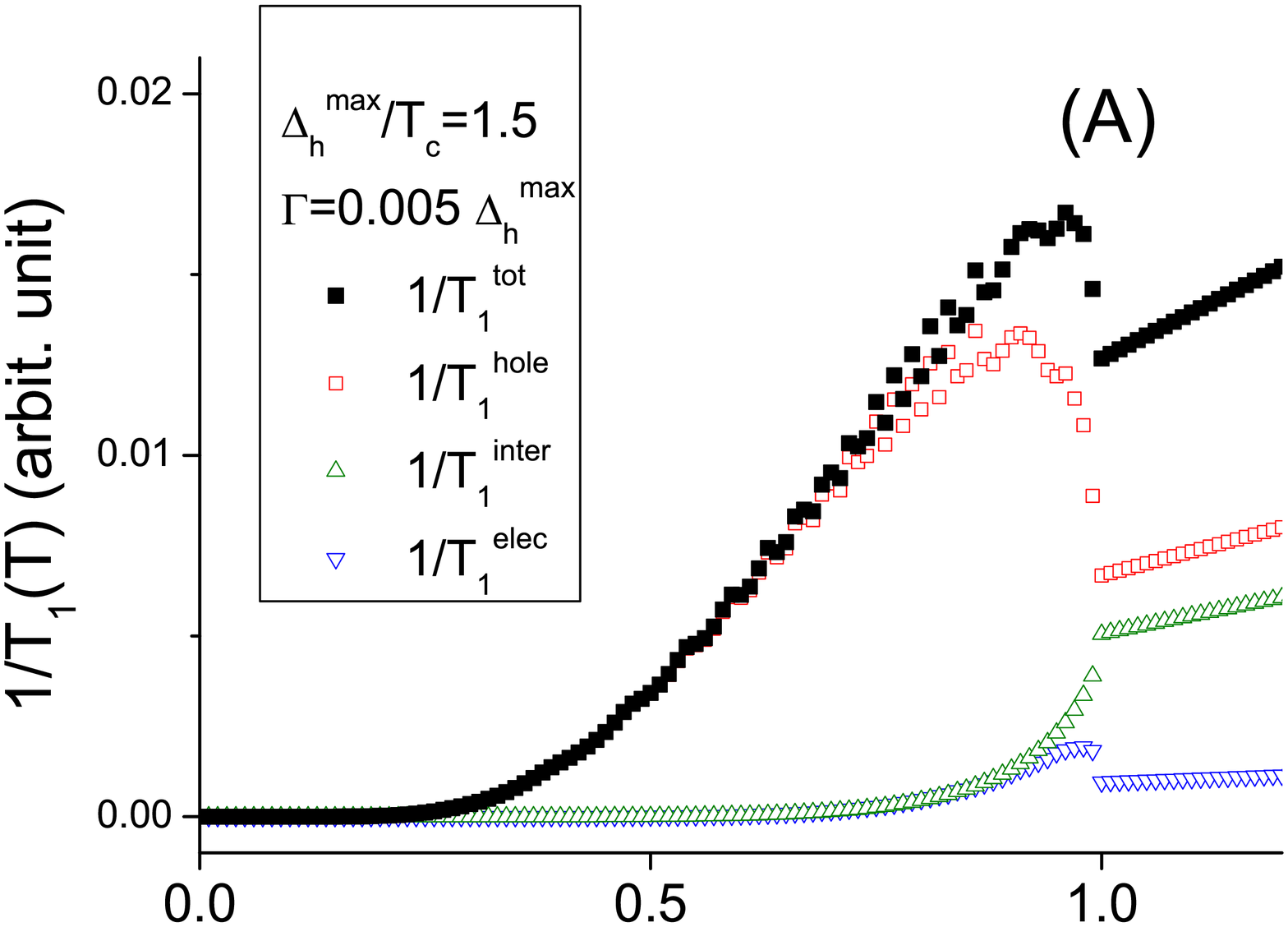}
\includegraphics[width=80mm]{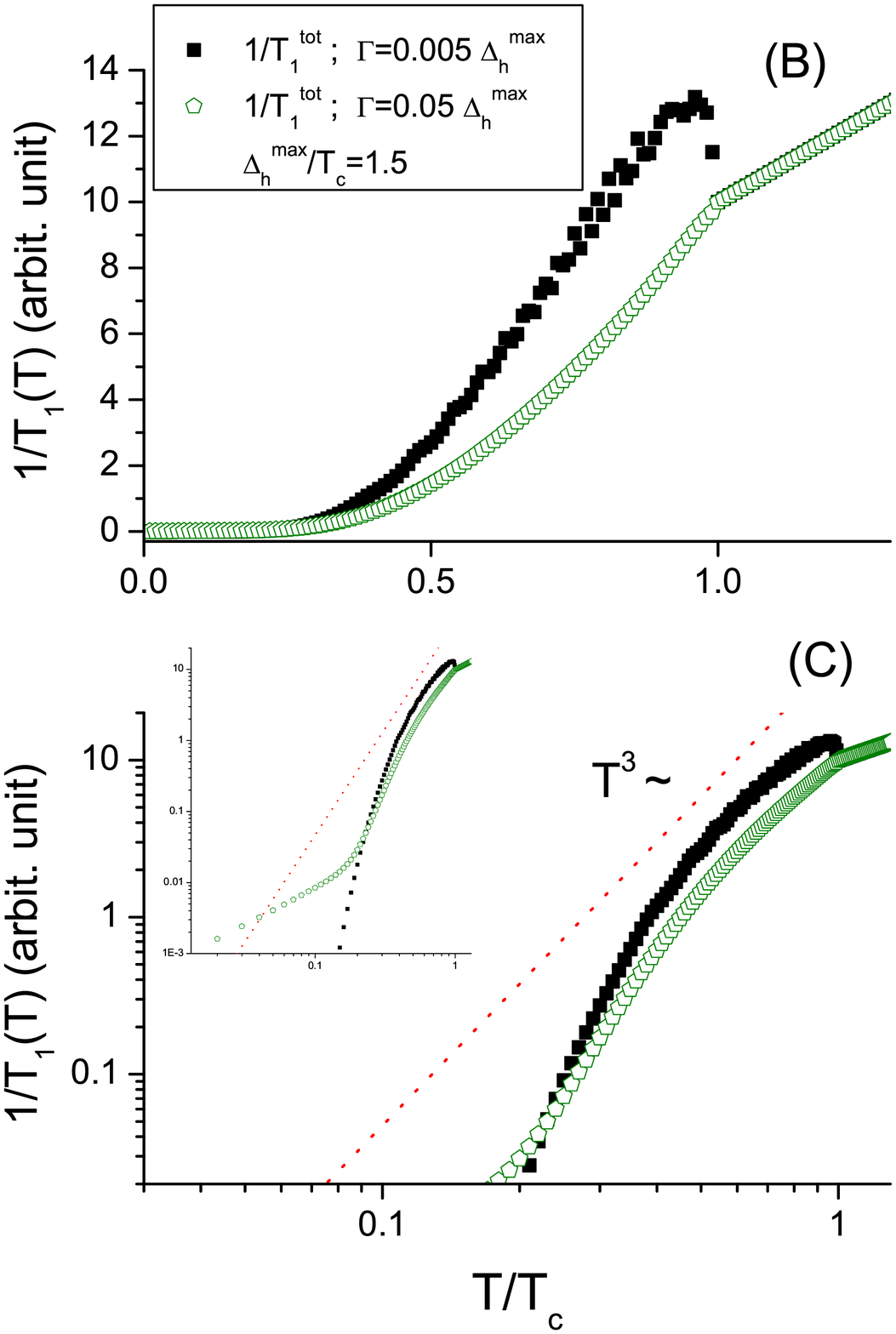}
\caption{(Color online) $1/T_1$(T) of the $\pm$s-wave gap with
$\Delta_h ^{max}/ T_c$ =1.5. (a) Separate term contributions of
Eq.(8): total (solid black square), hole band (open red square),
electron band (open blue triangle), and interband term (solid
green square). (b) Total $1/T_1$(T) without (solid black square)
and with (open green pentagon) damping. (c) The log-log plot of
(b). The inset is a wide view \label{fig3}}
\end{figure}

\begin{eqnarray}
\label{1 o T1} \frac{1}{T_1} &\sim&  - T \int_0 ^{\infty}
\frac{\partial f_{FD} (\omega)}{\partial \omega} \Biggl \lbrace
\Bigl[
 N_h ^2(0)\left\langle Re
   \frac{\omega}{\sqrt{\omega^2-\Delta_h ^2(k)}}
  \right\rangle_{k}^2
  \nonumber \\   &+& 2 N_h (0) N_e (0) \left\langle Re
     \frac{\omega}{\sqrt{\omega^2-\Delta_h ^2(k)}}
    \right\rangle_{k} \left\langle Re
     \frac{\omega}{\sqrt{\omega^2-\Delta_e ^2(k^{'})}}
    \right\rangle_{k^{'}}
 \nonumber \\   &+&  N_e ^2(0) \left\langle Re
     \frac{\omega}{\sqrt{\omega^2-\Delta_e ^2(k)}}
    \right\rangle_{k}^2
\Bigr] \nonumber \\
&+& \Bigl[
 N_h ^2(0)\left\langle Re
   \frac{\Delta_h(k)}{\sqrt{\omega^2-\Delta_h ^2(k)}}
  \right\rangle_{k}^2
  \nonumber \\   &+& 2 N_h (0) N_e (0) \left\langle Re
     \frac{\Delta_h(k)}{\sqrt{\omega^2-\Delta_h ^2(k)}}
    \right\rangle_{k} \left\langle Re
     \frac{\Delta_e(k^{'})}{\sqrt{\omega^2-\Delta_e ^2(k^{'})}}
    \right\rangle_{k^{'}}
 \nonumber \\   &+&  N_e ^2(0) \left\langle Re
     \frac{\Delta_e(k)}{\sqrt{\omega^2-\Delta_e ^2(k)}}
    \right\rangle_{k}^2
\Bigr]\Biggr \rbrace.
\end{eqnarray}

\noindent For the temperature dependence of the gaps
$\Delta_{h,e}(k,T)$, we use a phenomenological formula,
$\Delta_{h,e}(k,T)=\Delta_{h,e}(k,T=0) \tanh (\beta
\sqrt{T_{c}/T-1})$. By choosing the values of $\Delta _{h,e}
^{max} / T_c$, we can partially take into account of the strong
coupling superconductivity effect. $\beta$ is not a sensitive
parameter for final results; we take $\beta=1.74$ in this paper.

Fig.3(a) shows the contributions to the $1/T_1$ relaxation rate
from each terms of Eq.(8): the hole band, the electron band, and
the interband terms. It shows that the cancellation of the
coherence factors is not perfect in general unless the conditions
$N_h(0) = N_e(0)$ as well as $|\Delta_h (k)| = |\Delta_e (k)|$ are
fulfilled. Nevertheless, due to the large cancellation by the
interband coherence factor, the height of the coherence peak is
very much reduced [compare the total $1/T_1$ and the hole band
only $1/T_1$ in Fig.3(a)]. Small amount of impurities can easily
wash out this reduced coherence peak as shown in Fig.3(b); the
damping rate $\Gamma=0.05 \Delta_h ^{max}$ is enough to completely
kill the coherence peak. The subtle part is to fit the $\sim T^3$
power law below $T_c$. It requires to tune $R=\Delta _{h} ^{max} /
T_c$ ratio. In Fig.3, $R=1.5$ (automatically, it makes $\Delta
_{e} ^{max} / T_c \approx 3.75$ which is quite a large value) is
used for the best fit. Fig.3(c) shows that this pseudo-$T^3$
behavior is not extended to the very low-temperature region as in
the d-wave case because this $T^3$ behavior in the $\pm$s-wave gap
is not an intrinsic property of the lines of nodes. At low
temperatures, there appears the exponential drop inevitably due to
the full gaps, and then it finally reaches the impurity-induced
$T$-linear region because we added some amount of impurities to
kill the coherent peak. All these details put rather stringent
conditions to confirm the $\pm$s-wave gap state with experiments.

\begin{figure}
\noindent
\includegraphics[width=80mm]{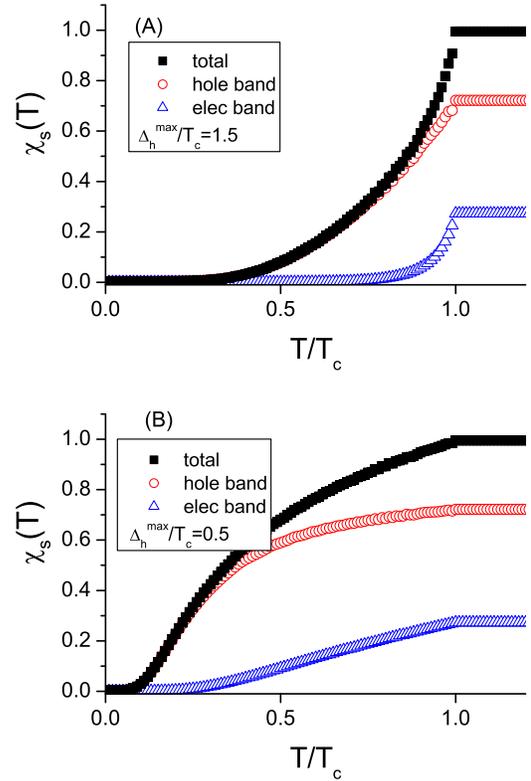}
\caption{(Color online)  Normalized Knight shift (uniform spin
susceptibility) of $\pm$s-wave gap. The total (solid black
square), hole band (open red circle), and electron band (open blue
triangle) contributions are shown separately. (a) $\Delta_h
^{max}/ T_c =1.5$ and (b)$\Delta_h ^{max}/ T_c$ =0.5.
\label{fig4}}
\end{figure}

\begin{figure}
\noindent
\includegraphics[width=80mm]{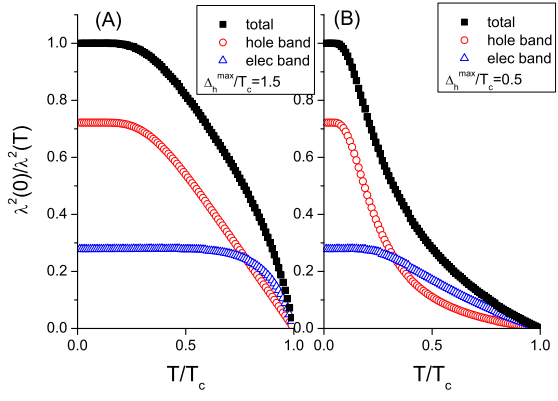}
\caption{(Color online)  Normalized superfluidity density
$\lambda^2 (0)/\lambda^2 (T)$ of $\pm$s-wave gap and its separate
contributions from the hole and electron bands. (a) $\Delta_h
^{max}/ T_c =1.5$ and (b)$\Delta_h ^{max}/ T_c$ =0.5.
\label{fig5}}
\end{figure}

Now we consider the Knight shift which is the measure of uniform
susceptibility in SC phase. Because it is a ${\bf q} \rightarrow
0$ probe, there is no interband contribution and the total Knight
shift is just sum of the contributions from each band as follows:

\begin{eqnarray}
\label{spin_susceptibility}  \chi_{S}(T) &\sim&  -\int_0 ^{\infty}
\frac{\partial f_{FD} (\omega)}{\partial \omega}
 \Big[ N_h(0) \left\langle Re
   \frac{\omega}{\sqrt{\omega^2-\Delta_h ^2(k)}}
  \right\rangle_{k}  \nonumber \\   &+&
N_e(0) \left\langle Re
   \frac{\omega}{\sqrt{\omega^2-\Delta_e ^2(k)}}
  \right\rangle_{k} \Big].
\end{eqnarray}

In Fig. 4, the normalized Knight shift (uniform spin
susceptibility) is plotted and it shows the typical flat behavior
of a s-wave gap at low temperatures.  The contributions from the
hole and electron bands show separately the feature of the larger
DOS with small gap and the smaller DOS with a larger gap. Fig.
4(A) is the results with $\Delta_h ^{max}/ T_c$ =1.5, the best
gap-$T_c$ ratio to fit $T^3$ behavior of $1/T_1$ in Fig.3. Fig4(B)
showed the results with $\Delta_h ^{max}/ T_c$ =0.5 to demonstrate
the convex shape of Knight shift  which was reported by Matano et
al.\cite{Curro} for Pr(FO)FeAs. Impurities does not change much of
this feature unlike in the case of d-wave gap.

Finally, we consider the penetration depth. The static response
function to the electromagnetic fields is the following:
\cite{leggett}

\begin{eqnarray}
\label{pene kernel}  K_{h,e} (q,T) &=&  2 \pi T \nonumber \\
& \times & \sum_n  \left\langle \hat{k}\| ^2
   \frac{\Delta_{h,e} ^2 (k)}{\sqrt{\omega_n ^2
   + \Delta_{h,e} ^2(k)} (\omega_n ^2 + \Delta_{h,e} ^2(k) + \alpha^2)}
  \right\rangle_{k}.
\end{eqnarray}

The ${\bf q} = 0$ limit of this kernel $K(q=0,T)$ is directly
proportional to the superfluidity density or 1/$\lambda_{L} ^2
(T)$ in the London limit. For our two band model, total kernel is
the sum of $K_h(q,T)$ and $K_e(q,T)$ with the proper weighting
factor proportional to the DOS N$_{h,e} (0)$ of each band and
there is no interband screening current. $\alpha=(v_F /2)\vec{\bf
q} \hat {{\bf k}}$ is the non-local parameter and can be rewritten
in more convenient form as $\alpha = (\frac{\xi_0}{\lambda_0})
\tilde{\bf q} \hat {{\bf k}}$. $\xi_0 \approx v_F/ \Delta^{max} $
and $\lambda_0$ are the coherence length and the penetration depth
at zero temperature, respectively. Apparently if
$(\frac{\xi_0}{\lambda_0})=\alpha_0$ is small compared to 1, the
non-local effect becomes negligible. A typical value of $\alpha_0$
for YBCO was estimated about 0.01, for example \cite{leggett}. For
the Fe-based superconductors, we believe that $\alpha_0$ is not
much larger than the values of the high-$T_c$ cuprates.  Also for
a s-wave case, the non-local effect does not change much of the
temperature dependence of $1/\lambda^2 (T)$ except the overall
magnitude. Therefore, we take $\alpha_0=0.0$ for the calculations
of the penetration depth in the $\pm$s-wave case. However, this
effect can induce an important modifications in the d-wave case,
which will be discussed in Sec.IIIB.

Fig.5 shows the normalized  superfluidity density $\lambda^2
(0)/\lambda^2 (T)$ and separate contributions from the hole and
electron bands.  The exponentially flat region appears at low
temperatures due to the full gap opening, which is consistent with
recent experiments.\cite{pene} Relatively narrower region of the
flat part (for T $< 0.2 T_c$) compared to the ordinary s-wave gap
is another feature due to the smaller gap with the larger DOS of
the $\pm$s-wave gap SC. A subtle part here is to fit the high-
temperature region (0.3 T$_c < T < T_c$). With $\Delta_h ^{max}/
T_c$ =1.5 (the same  value used for the $1/T_1$ fit), this part
becomes too convex [Fig.5(A)] in comparison to the experiments.
\cite{pene} A smaller gap-$T_c$ ratio can make it concave as shown
in Fig.5(B) (with $\Delta_h ^{max}/ T_c$ =0.5); this concave
feature was recently observed by Martin et al. \cite{pene}

\begin{figure}
\noindent
\includegraphics[width=80mm]{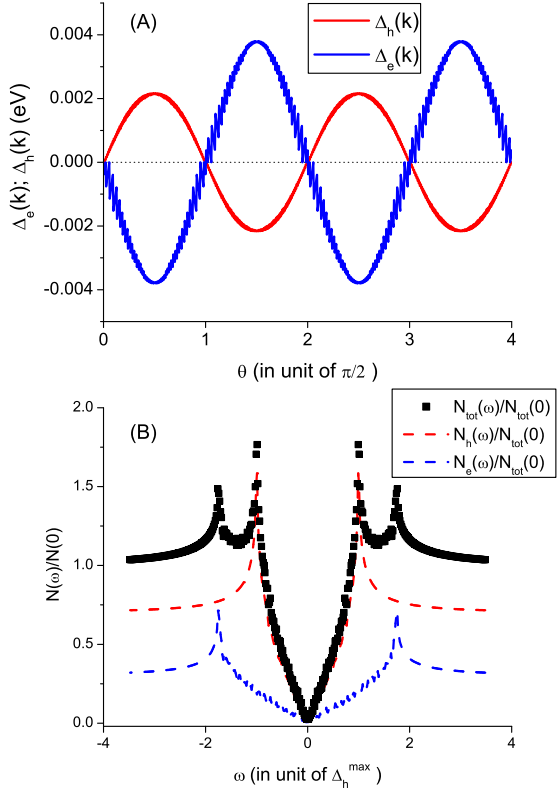}
\caption{(Color online) (a) Double d-wave gap solutions $\Delta_h
(k)$ and $\Delta_e (k)$. (b) Normalized DOS of the hole band
N$_{h} (\omega)$ (red dotted line), electron band N$_{e} (\omega)$
(blue dotted line), and the total N$_{tot} (\omega)$ (black
squares). \label{fig6}}
\end{figure}

In summary, the $\pm$s-wave gap state provides the most consistent
descriptions for the penetration depth experiments. However, it
explains $1/T_1$ only for a limited temperature range, even with a
fine tuning of $\Delta_{h,e}/T_c$ ratio and impurities.  Knight
shift of any shape can be fit with two band parameter (this is
also true with the double d-wave gap). The tunneling DOS does not
provide a decisive conclusion.

\begin{figure}
\noindent
\includegraphics[width=80mm]{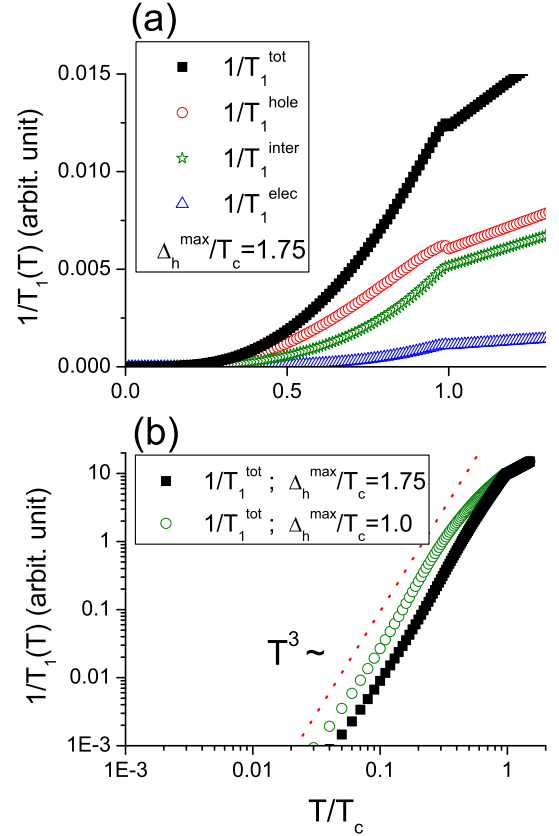}
\caption{(Color online) $1/T_1$(T) of the double d-wave gap. (a)
Each term contributions of Eq.(8) : total (solid black square),
hole band (open red square), electron band (open blue triangle),
and interband term (open green triangle). (b)Log-log plot of total
$1/T_1$(T) for $\Delta_h ^{max}/ T_{c}=1.75$ (black square) and
1.0 (green circle).  \label{fig7}}
\end{figure}

\subsection {Double d-wave gap}

In Fig.6, the gap solution and the corresponding DOS of the double
d-wave gap are shown. As mentioned, our model bands have
$N_h(0)=2.6 N_e(0)$, and consequently gap in the hole band
$\Delta_h ^{max} \approx 2 meV$ is smaller than the one of the
electron band $\Delta_e ^{max} \approx 4 meV$. The sizes of the
maximum gaps are $\sim$ 5 times smaller than the $\pm$s-wave gap
solutions. Therefore, the double d-wave gap solution is not the
best SC state for our phenomenological model with an
antiferromagnetic pairing interaction. This result is in agreement
with other theoretical studies \cite{Mazin1,DHLee,Eremin}. This
conclusion may change with the correlation length of the  AFM
fluctuations, the sizes of the FS of the hole and electron bands,
etc. But we numerically found that $\pm$s-wave gap solution is
favored compared to the double d-wave gap solution for most cases.
As discussed in Sec.II, however, the screened Coulomb interaction
may change this conclusion. The detailed studies about this issue
will be discussed in a separate paper.

To complete the comparisons, we calculated the same SC properties
of the double d-wave gap state. The separate and total DOSs shown
in Fig.6(b) demonstrates the main features of the double d-wave
gap: the large DOS band with a small gap and the small DOS band
with a large gap. This result shows a similar feature of the
tunneling DOS measurement by Wang et al. \cite{tunneling} except
the ZBCP, which does not show up in our simple DOS calculation but
should appear when the tunneling conductivity is properly
calculated with Andreev scattering process.

We consider the nuclear spin-lattice relaxation rate $1/T_1$. As
in the case of $\pm$s-wave gap, there are three contributions for
the total $1/T_1$ relaxation rate: hole band, electron band, and
interband scattering terms. The formula is the same as Eq.(8) but
the last three terms should be dropped because the FS average of
$\Delta_{h,e} (k)$ vanishes in this case.  We use the same form of
temperature dependent gap function as
$\Delta_{h,e}(k,T)=\Delta_{h,e}(k,T=0) \tanh (1.74
\sqrt{T_{c}/T-1})$. In Fig.7, $R=\Delta_h ^{max}/ T_{c}$=1.75 is
used for the best $T^3$ fit below $T_c$. However, in the double
d-wave gap, $R= 1.5 - 2.5 $ provide reasonably good fits, showing
a more tolerance than the $\pm$s-wave gap state.

\begin{figure}
\noindent
\includegraphics[width=80mm]{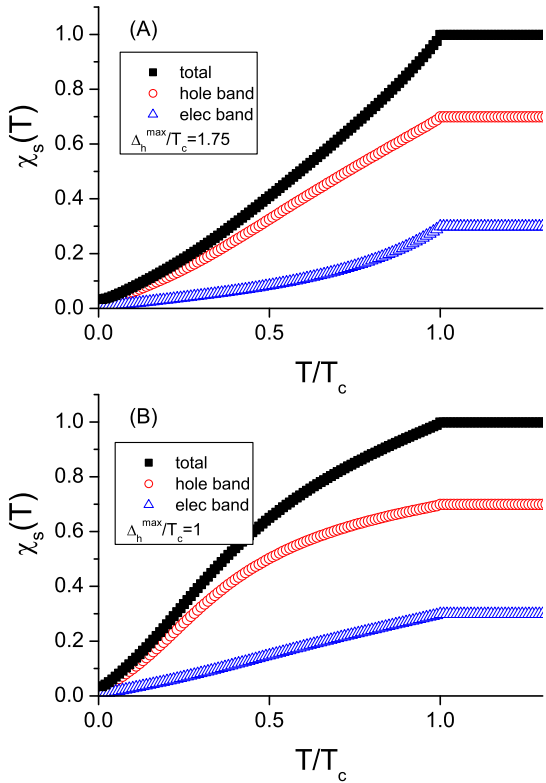}
\caption{(Color online) Normalized Knight shift (uniform spin
susceptibility) of the double d-wave gap: the total (solid black
square), the hole band (open red circle), and electron band (open
blue triangle) contributions are shown separately. (a) $\Delta_h
^{max}/ T_c =1.75$ and (b) $\Delta_h ^{max}/ T_c$ =1.0
\label{fig8}}
\end{figure}

Fig.7(a) shows the separate contributions from each channel
together with the total contribution. As in the $\pm$s-wave case,
$1/T_{1,h}$ provides the largest contribution and $1/T_{1,e}$
provides the smallest contribution. There is also the interband
term $1/T_{1}  ^{inter}$. In contrast to the $\pm$s-wave case, all
three terms display a similar temperature dependence and no
coherence peaks. Fig.7(b) shows the same $1/T_{1} ^{total}$ (black
squares) in log-log plot. The overall features of $1/T_1$ are the
ones of the typical d-wave SC state: no coherence peak near $T_c$
and $\sim T^3 $ below $T_c$, and consistent with the current NMR
experiments \cite{Curro,T1}. At very low temperatures, T-linear
behavior starts to appear due to a small damping for the numerical
calculations ($\Gamma=0.005 \Delta_h ^{max}$). For comparison, we
also show $1/T_{1} ^{total}$ (open green circles) with $\Delta_h
^{max}/ T_c$ =1.0, a smaller gap-$T_c$ ratio; it exhibits a
substantial convex part below $T_c$ and then starts displaying the
universal $T^3$ behavior before entering the impurity dominating
region.

In Fig.8, we show the result of the uniform spin susceptibility
which is measured as Knight shift. Fig.8(a) shows the results with
$\Delta_h ^{max}/ T_{c}=1.75$. The hole band contribution is
dominant as in $1/T_1$ and the electron band contribution show the
steeper drop just below $T_c$ because of the larger gap-$T_c$
ratio $\Delta_e ^{max}/ T_{c}\approx 3.5$. The overall behavior of
the total $\chi_s (T)$ below $T_c$ shows a typical d-wave behavior
such as $T$-linear at low temperatures.
Fig.8(b) shows the results with $\Delta_h ^{max}/ T_{c}=1.0$. A
smaller gap-$T_c$ ratio makes the Knight shift convex as in the
$\pm$s-wave case and observed by Matano et al.\cite{Curro} for
Pr(FO)FeAS. This result demonstrates that this convex behavior of
Knight shift is irrelevant to the gap symmetry but a generic
feature of the two-gap (or multigap) SC. But it reveals that the
gap-$T_c$ ratio $\Delta^{max} (0)/T_c$ is much smaller than the
standard BCS value, where $\Delta^{max} (0)$ refers to the gap of
the band with largest DOS.

Now we calculate the penetration depth. As we discussed in Sec.II,
most of experiments, up to now, report a flat temperature
dependence of $\lambda (T)$ at low-temperature region and suggest
a fully gapped SC state \cite{pene}. A naive double d-wave gap
state has no chance to explain this flat behavior at low
temperatures. Therefore, we consider a non-local effect of the
electromagnetic response of the double d-wave gap superconductor
as a possible cause to modify the typical temperature dependence.
In order to include the effect of the non-local electrodynamics,
we use the fully ${\bf q}-$dependent kernel $K_{h,e}(q,T)$
[Eq.(10)] and put it into the integral formula for $\lambda(T)$
with the specular boundary condition,

\begin{eqnarray}
\label{pene depth} \frac{\lambda_{spec} (T)}{\lambda_0} &=&
\frac{2}{\pi} \int_{0} ^{\infty} \frac{d \tilde{q}}{\tilde{q}^2 +
N_h(0) \tilde{K}_h (q,T)+ N_e(0)\tilde{K}_e (q,T)}.
\end{eqnarray}

\noindent where $\tilde{K}_{h, e} (q,T)$ are the normalized
kernels as $\tilde{K}_{h, e} (0,0)=1$ and $\tilde{q}=q \lambda_0$
is a dimensionless momentum. The results with a diffusive boundary
condition are qualitatively the same; therefore, they will not be
discussed. For the non-local parameter $\alpha_0$, we think it to
be much smaller than 1 for the Fe-based superconductors, but here
we take it as a free parameter and see how large value is needed
to fit experimental data.

Fig.9 and Fig.10 show the normalized total superfluidity density
$\lambda^2 (0)/\lambda^2 (T)$ and separate contributions from the
hole and electron bands for the double d-wave gap state. Fig.9
used $\Delta_h ^{max} /T_c$=1.75 and Fig.10 used $\Delta_h ^{max}
/T_c$=1.0. In each figure, panel (A) is a local limit
($\alpha=0.0$) and the panel (B) is a non-local limit
($\alpha=0.5$).
The local cases display the typical d-wave behavior at low
temperatures, i.e., the linear decrease in $T$ from $T=0$. The
extreme non-local cases ($\alpha_0=\frac{\xi_0}{\lambda_0}=0.5$)
introduce a substantial round-off ($\sim T^2$) region at low
temperatures, which is, however, not an exponentially flat
behavior as the recent experiments claim. Further, even a rough
fitting requires an unreasonably large non-local parameter
$\alpha_0$.

\begin{figure}
\noindent
\includegraphics[width=80mm]{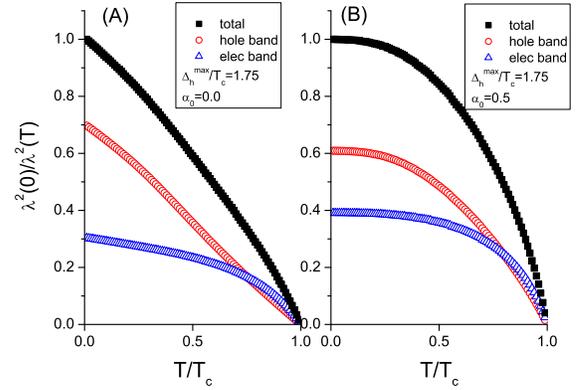}
\caption{(Color online)  Normalized superfluidity density
$\lambda^2 (0)/\lambda^2 (T)$ of double d-wave gap and its
separate contributions from the hole and electron bands with
$\Delta_h ^{max}/ T_c =1.75$. (a)
$\alpha_0=\frac{\xi_0}{\lambda_0}=0.0$ and (b) $\alpha_0$ =0.5.
\label{fig9}}
\end{figure}

\begin{figure}
\noindent
\includegraphics[width=80mm]{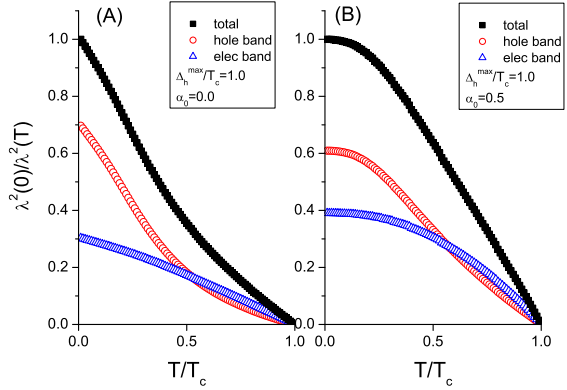}
\caption{(Color online)  Normalized superfluidity density
$\lambda^2 (0)/\lambda^2 (T)$ of double d-wave gap and their
separate contributions from the hole and electron bands with
$\Delta_h ^{max}/ T_c =1.0$. (a)
$\alpha_0=\frac{\xi_0}{\lambda_0}=0.0$; (b) $\alpha_0$ =0.5.
\label{fig10}}
\end{figure}

In summary, the double d-wave gap state can provide consistent
descriptions for tunneling DOS, $1/T_1$, and Knight shift.
However, there is an intrinsic difficulty to explain the flat
behavior of the penetration depth at low temperatures. Also, in
our model with an AFM mediated pairing interaction only, the
double d-wave gap solution is energetically less favored than the
$\pm$s-wave gap solution.

\section{Conclusion}

We demonstrated that a minimal model with a phenomenological
pairing interaction of the AFM spin fluctuations can allow both
the $\pm$s-wave gap and the double d-wave gap solutions with the
realistic bands of the Fe-based SC compounds. With the same
parameters, the $\pm$s-wave gap solution is energetically more
favorable by a factor of $\sim$5 times, so that it has a better
chance to be realized in the Fe-based SC compounds.

In both cases, we found that the approximate relation  $\Delta_{h}
^{max} N_h \approx \Delta_{e} ^{max} N_e$ holds because it is a
generic feature of the two gap SC when an interband pair
scattering is the dominant pairing interaction. This relation
appears for all SC properties in subtle way, which modifies the
value of $\Delta(0)/T_c$ and other SC properties in unorthodox
way. Numerically solving the coupled gap equations for the two
bands, we found the detailed structure of the gap functions
$\Delta_{h,e}(k)$, which showed an anisotropy ($\sim$ 20 $\%$) of
the $\pm$s-wave gaps. We also calculated the key SC properties,
for both gap states, such as tunneling DOS, $1/T_1$, Knight shift,
and penetration depth and discussed them in comparison with
experiments. When we calculated these quantities, we paid a
special attention to the interband coherence factor which is a
unique feature of multigap SC. This interband coherence factor
particularly produced an important modification to the $1/T_1$
relaxation rate of the $\pm$s-wave gap state.

The $\pm$s-wave gap state provides the most consistent
descriptions for the penetration depth experiments: the flat low-
temperature behavior.\cite{pene} Besides the low-temperature
behavior, the high-temperature (0.3 $T_c < T < T_c$) behavior --
due to a large difference of the gap sizes $\Delta_h$ and
$\Delta_e$ and their corresponding DOS $N_{h,e}$ -- can be either
concave or convex. However, $1/T_1$ experiments can only be fitted
for a limited temperature range, even with a fine tuning of
$\Delta_{h,e}/T_c$ ratio and impurities. The $\pm$s-wave gap state
is not inconsistent with the Knight shift and the tunneling DOS
data, but overall does not provide any decisive merit in
comparison with the double d-wave gap.

The double d-wave gap state, although it is energetically less
favorable in our model unless additional interactions are added,
provides the best fit to the $1/T_1$ experiments. However, it has
a difficulty to explain the penetration depth experiments for the
low temperature flat behavior. It requires an unreasonably large
non-local effect to fit the low temperature part; it is still not
exponentially  flat but only $\sim T^2$. If this low temperature
part of $\lambda(T)$ is, indeed, confirmed to be exponentially
flat, the double d-wave gap state should be ruled out. Tunneling
DOS and Knight shift can be fit with the double d-wave gap state
as much as with the $\pm$s-wave gap state.

In conclusion, quantitative calculations, carried out in this
paper, with the two most promising SC gap states can serve as
guidelines for sorting out the possible pairing states of the
Fe-based SC in comparison with the current and future experiments.
For that, very low temperature measurements and systematic studies
with the amount of impurities will provide decisive information to
determine the correct gap symmetry.

{\it Note added --} Recently, we have known that similar studies
of $1/T_1$ for the $\pm$s-wave state were carried out by two
groups \cite{recent} where only the interband scattering process
was analyzed and by another recent paper \cite{Parish} where both
the interband and intraband processes were considered as in our
paper.

\section*{Acknowledgement}
This work was supported by the KOSEF through the Grants No.
KRF-2007-070-C00044, and No. KRF-2007-521-C00081.

\appendix
\numberwithin{figure}{section}
\numberwithin{equation}{section}

\section{Relation between $\Delta_h / \Delta_e$ and $N_h / N_e$}

\begin{figure}
\noindent
\includegraphics[width=90mm]{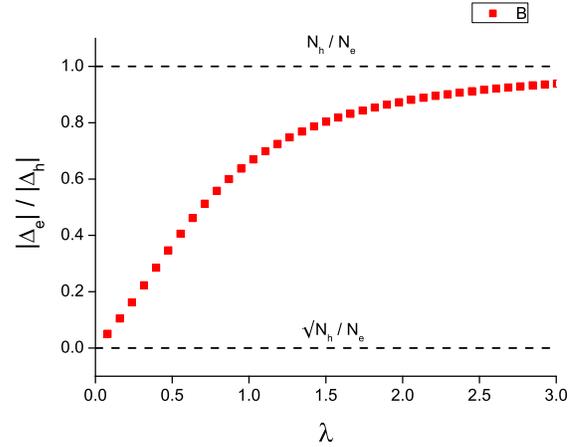}
\caption{(Color online)  The normalized gap ratio $|\Delta_e | /
|\Delta_h |$ vs the dimensionless coupling constant $\lambda=
\sqrt{V_{he} V_{eh} N_h N_e}$. The bottom baseline is $\sqrt{N_h /
N_e}$ and the top baseline is $N_h / N_e$. \label{Fig1}}
\end{figure}

In the main text, we claimed the approximate relation $N_h
\Delta_h \approx N_e \Delta_e$ as a generic feature of the two-
band model with a dominant interband interaction. This kind of
relation will have direct and important implications to the
experimental observations. However it is pointed out by Mazin
\cite{Mazin-app} that this is not a rigorous identity in general.
In this appendix, we clarify the degree of the validity of this
relation.

Here we consider the $\pm$s-wave gap state only. Assuming constant
gaps, $\Delta_{h} (k) =\Delta_{h}$ and $\Delta_{e} (k)
=-\Delta_{e}$, and only the interband interaction, the coupled gap
equations (5) and (6) are simplified as

\begin{eqnarray}
\Delta_h &=&  V_{he} \Delta_e \chi_e (T, \Delta_e, \omega_{AFM})  \\
\Delta_e &=&  V_{eh} \Delta_h \chi_h (T, \Delta_h, \omega_{AFM})
\end{eqnarray}

\noindent where $\chi_h$ and $\chi_e$ are defined with Eq.(7).

First, when $T=T_c$, the above equations can be written as

\begin{eqnarray}
\Delta_h = V_{he} N_e \Delta_e \times const  \\
\Delta_e = V_{eh} N_h \Delta_h \times const
\end{eqnarray}

\noindent where $const = \int_0 ^{\omega_{AFM}} d \xi
\frac{\tanh{\frac{\xi}{2 T_c}}}{\xi} \approx \log{1.14
\omega_{AFM} /T_c}$. Because of $V_{he}=V_{eh}$, we immediately
obtain the following relation from the above equations.

\begin{equation}
\frac{\Delta_e}{\Delta_h} = \sqrt{\frac{N_h}{N_e}}~~; ~~~{\rm
when} ~T=T_c.
\end{equation}

Next, when $T=0$, Eqs. (A.1) and (A.2) are written as

\begin{eqnarray}
\Delta_h &=& V_{he} N_e \Delta_e \log (\frac{\omega_{AFM}+\sqrt{\omega_{AFM}^2+\Delta_e ^2}}{\Delta_e}) \\
\Delta_e &=& V_{eh} N_h \Delta_h  \log
(\frac{\omega_{AFM}+\sqrt{\omega_{AFM}^2+\Delta_h ^2}}{\Delta_h})
\end{eqnarray}

\noindent In general, these equations do not yield a simple
algebraic relation between $\Delta_e / \Delta_h$ and $N_h / N_e$,
but we can obtain the simple relations for the limiting cases.
First, for the extreme weak coupling limit, ie., when
$\Delta_{h,e} \ll \omega_{AFM} $, the two logarithmic terms become
asymptotically equal as $\log (2 \omega_{AFM}/ \Delta_e) \approx
\log (2 \omega_{AFM}/ \Delta_h)$,  and we obtain the same relation
as the $T=T_c$ case Eq.(A.5).
On the other hand, for strong-coupling limit, ie., when
$\Delta_{h,e} \gg \omega_{AFM} $ (which is certainly an unphysical
limit), $\log ([\omega_{AFM}+\sqrt{\omega_{AFM}^2+\Delta_{h,e}
^2}]/\Delta_{h,e}) \approx \omega_{AFM}/\Delta_{h,e}$ and we
obtain the relation,

\begin{equation}
\frac{\Delta_e}{\Delta_h} = \frac{N_h}{N_e}~~; ~~~{\rm when}
~~T=0~~ {\rm and}~~ \Delta_{h,e} \gg \omega_{AFM}.
\end{equation}

Having found the results of the two limiting cases, we can guess
that the gap ratio $\Delta_e / \Delta_h$ should be in between
these two limiting ratios. For example, we can attempt an
expansion with $x=\log{[N_h/N_e]}$ starting from the weak coupling
limit \cite{Mazin-app}, and we obtain, in the first order of $x$,

\begin{equation}
\frac{\Delta_e}{\Delta_h} \approx \sqrt{\frac{N_h}{N_e}}
(1+\frac{\log{[N_h/N_e]}}{4} \lambda +  \ldots)
\end{equation}

\noindent where $\lambda= \sqrt{V_{he} V_{eh} N_h N_e}$ is a
dimensionless coupling constant. For practical use, we numerically
solve Eqs. (A.1) and Eq.(A.2) and plot the ratio $\Delta_e /
\Delta_h$ as a function of $\lambda$ in Fig.A.1. The ratio
$\Delta_e / \Delta_h$ becomes a universal curve when it is
normalized by the distance between $\sqrt{\frac{N_h}{N_e}}$ and
$\frac{N_h}{N_e}$. The result indeed shows that when $\lambda
\approx 1$, it is in between two limiting ratios,
$\sqrt{\frac{N_h}{N_e}}$ and $\frac{N_h}{N_e}$ as we expected from
the above analysis.

In reality, there exist two complications. First, the intraband
couplings $V_{hh}$ and $V_{ee}$ need to be included. A little
analysis of Eqs. (5) and (6) as well as of numerical results
reveals that this effect always enhances the gap ratio toward the
limit $\frac{N_h}{N_e}$. Another complication arises from the fact
that there are more than two bands in real materials
\cite{Singh,Haule,Mazin1,Mazin2,Cao,bands}. Applying the results
of the above analysis, we can suggest the following approximate
relations. First, we classify the bands of the real materials into
two groups: the hole bands around $\Gamma$ point and the electron
bands around $M$ point, respectively. Then in the strong-coupling
limit,

\begin{equation}
\sum_i \Delta_{h,i} N_{h,i} \approx \sum_i \Delta_{e,i} N_{e,i},
\end{equation}

\noindent and in the extreme weak-coupling limit,

\begin{equation}
\sum_i \Delta_{h,i} \sqrt{N_{h,i}} \approx \sum_i \Delta_{e,i}
\sqrt{N_{e,i}}.
\end{equation}

Considering several uncertainties in reality,  Eq.(A.10) can serve
as a practical rule of the thumb.


\begin{references}
\bibitem{Mazin-app} This appendix section is motivated by the communication with I.I. Mazin and a
large part of the analysis is directly suggested by him.
\end{references}
\end{document}